\documentclass[useAMS,usenatbib]{mn2e}

\usepackage{graphicx}
\usepackage{txfonts}
\usepackage{epsfig}
\usepackage{tabularx}
\usepackage{subfig}

\title[The formation of blue hook stars in globular cluster]
{A possible formation channel for blue hook stars in globular cluster}
\author[Lei et al.]{Zhenxin Lei$^{1,2}$\thanks{E-mail:
lzx2008@ynao.ac.cn}, Xuemei Chen$^{3}$, Fenghui Zhang$^{2,4}$ and Zhanwen Han$^{2,4}$\\
$^{1}$Department of Science, Shaoyang University, Shaoyang 422000, China\\
$^{2}$Key Laboratory for the Structure and Evolution of Celestial Objects,
Chinese Academy of Sciences,Kunming 650011,China\\
$^{3}$Department of Electrical Engineering, Shaoyang University, Shaoyang 422000, China\\
$^{4}$Yunnan Observatory, Chinese Academy of Sciences, Kunming 650011, China}

\begin{document}

\date{Accepted ; Received ; in original form}

\pagerange{\pageref{firstpage}--\pageref{lastpage}} \pubyear{2015}

\maketitle

\label{firstpage}

\begin{abstract}

The formation mechanism for blue hook (BHk) stars in
globular clusters (GCs) is still unclear. Following
one of the possible scenario, named late hot flash scenario,
we proposed that tidally enhanced stellar wind in binary
evolution may provide the huge mass loss on the red giant
branch (RGB) and produce BHk stars. Employing the
detailed stellar evolution code, Modules for
Experiments in Stellar Astrophysics (MESA), we
investigated the contributions of tidally enhanced stellar wind
as a possible formation channel for BHk stars in GCs. 
We evolved the primary stars with different initial 
orbital periods using the binary module in MESA  
(version 6208) from zero age main-sequence (ZAMS) to post horizontal branch 
(HB) stage, and obtained their evolution parameters which are compared with 
the observation.
The results are consistent with observation in the 
color-magnitude diagram (CMD) and 
the $\rm{log}\it{g}-\it{T}\rm_{eff}$ plane for
NGC 2808, which is an example GC hosting BHk stars.
However, the helium abundance in the surface for our models
is higher than the one obtained in BHk stars. This
discrepancy between our models and observation is
possibly due to the fact that gravitational settling and
radiative levitation which are common processes in
hot HB stars are not considered in the models as well
as the fact that the flash mixing efficiency may
be overestimated in the calculations. Our
results suggested that tidally enhanced stellar
wind in binary evolution is able to naturally provide the
huge mass loss on the RGB needed for
late hot flash scenario and it is
a possible and reasonable formation channel
for BHk stars in GCs.

\end{abstract}


\section{Introduction}

For a long time, globular clusters (GCs) have been known as the
fundamental laboratory for studying stellar structure and evolution,
since they are considered to be composed of simple stellar populations
(SSP) in which all member stars are formed at the same time with similar
chemical abundance. However, this point of view is challenged by recent
observations both from photometry and spectroscopy.
For example, the splitting of main-sequence 
(MS) has been found in some massive GCs, such as $\omega$ Cen
(Bedin et al. 2004), NGC 2808 (Piotto et al. 2007) etc.
Moreover, star to star variation of light elements
(e.g., Na-O, Mg-Al anticorrelation) are found in
most of the studied GCs (Carretta et al. 2009; Gratton et al. 2012).
Both the splitting of MS and light elements variation
in GCs are considered to be possibly related to the
self-enrichment scenario with helium enhancement 
(D'Antona \& Caloi 2008; Milone 2015;
but also see Jiang et al. 2014 for an alternative solution).
In this scenario, the second generation
stars are helium enhanced, since they are formed in the material polluted by
the ejecta from either massive asymptotic giant branch (AGB) stars
(Ventura et al. 2001, 2002) or fast rotating massive stars (Decressin et al. 2007)
of the first generation. These second generation stars will occupy
a bluer position on MS and horizontal branch (HB) stage than normal
stars. In some GCs, the helium abundance for the second generation stars
may reach as high as $Y$ = 0.4 (Busso et al. 2007).

Multiple populations in GCs are also found in late stellar evolution stage,
such as HB. Recently, Na-O anticorrelation
is also discovered in HB stars for many GCs (Gratton et al. 2013, 2014).
Marino et al. (2014) analyzed
nearly 100 HB stars with different temperatures in NGC 2808 and
confirmed that some blue HB stars in this GC present higher
helium abundance than primordial content by
$\triangle Y$ = 0.09 $\pm$ 0.01. This result
provides direct observation constraints on the typical
second-parameter problem in GCs which is a long-term puzzle
in stellar evolution (see Catelan 2009 for a recent review).
In 1960, Sandage \& Wallerstein found that the
position of HB stars on color-magnitude diagram (CMD),
which is defined as HB morphology, is mainly determined by
metallicity of GCs. However, other parameters are also needed
to explain the diversity of HB morphology in GCs, such as
age (Lee et al. 1994; Dotter et al. 2010; Gratton et al. 2010),
helium enhancement due to self-enrichment
(D'Antona et al. 2002; D'Antona \& Caloi 2004; Milone et al. 2014),
stellar rotation (Sweigart et al. 1997),
binaries (Lei et al 2013a, 2013b, 2014; Han et al. 2012; Han \& Lei 2014), 
dynamics (Pasquto et al. 2013, 2014), etc.

More complicated situations arise from the discovery
for a special population of hot HB stars in some massive GCs,
which are called the blue hook (BHk) stars
(Whitney et al. 1998; D'Cruz et al. 2000).
These stars occupy a very blue position on the HB but with fainter
luminosity than the normal extreme horizontal branch (EHB) stars.
Due to its high effective temperature
(e.g., $T\rm_{eff}$ $\geq$ 32000 K; Moni Bidin et al. 2012) and faint luminosity,
BHk star can not be explained by canonical stellar evolution theory,
and its formation mechanism is still unclear.
There are several scenarios having been proposed
to explain the formation of BHk stars.
Brown et al. (2001) suggested that late hot flash scenario
is one of the possible explanations.
In this scenario, the helium core flash of
low mass stars may take place when the star is
descending the white dwarf cooling
curve if their progenitors undergo unusually huge mass loss
on the red giant branch (RGB) stage. 
During the helium core flash,
hydrogen in the envelope can be mixed into
interior and burn into helium at higher temperature.
Thus, these stars show helium and carbon enhancement
in the surface and occupy a bluer and
fainter position than canonical EHB stars in the CMD of GCs.
However, the physical mechanism for the huge mass loss on the
RGB in late hot flash scenario is still unclear.
On the other hand, Lee (2005) suggested that
the BHk stars in $\omega$ Cen and NGC 2808
are likely the progeny of a minority population of bluer and fainter
MS stars which could be reproduced by
assuming a large variations of primordial helium abundance
(e.g., $\bigtriangleup Y \approx$ 0.15), while
D'Antona et al. (2010) proposed that if
the blue MS stars found in $\omega$ Cen are
the progenitors of the BHk stars in this cluster,
they need to suffer extra mixing processes on the RGB and
increase their surface helium abundance up to $Y\approx$ 0.8. 

In this paper, we followed the late hot flash scenario and
suggested that tidally enhanced stellar wind in binary
evolution is a possible formation channel for BHk stars in GCs.
In binary evolution, the mass loss of a red giant
primary star could be largely enhanced by its companion star
(Tout \& Eggleton 1988), especially when the primary star
nearly fills its Roche lobe. In this scenario,
the mass loss of the red giant primary star depends on
orbital period of the binary system.
Some of the primary stars may lost nearly the whole
envelope before Roche lobe overflow (RLOF) is taking
place due to a relative shorter orbital period.
They evolve off the RGB and may experience helium
core flash  when descending the white
dwarf cooling curve (Castellani \&
Castellani 1993; D'Cruz et al. 1996; Brown et al. 2001, 2010, 2012).
After the helium core flash, the star burns helium stably in
its core and locates on the BHk regions in CMD of GCs.
Employing the detailed stellar evolution code, Modules for
Experiments in Stellar Astrophysics (MESA; Paxton et al. 2011, 2013),
we investigated the contributions of tidally enhanced stellar
wind in binary evolution on the formation of BHk stars in GCs.
In Section 2, we introduce the code and models used in this paper.
The results and comparison with observations are given in
Section 3. Finally, a discussion and conclusion are presented
in Section 4 and 5 respectively.

\section[]{Model construction}

The tidally enhanced stellar wind was firstly
suggested by Tout \& Eggleton (1988) to explain
the mass inversion phenomenon (e.g., more evolved
star has a low mass) found in some binaries of
RS CVn type (Popper \& Ulrich 1977; Popper 1980).
They suggested that the secondary star may tidally
enhance the stellar wind of the red giant primary star.
By considering tidally enhanced stellar wind in
binary evolution, they successfully explained the mass of
Z Her which is a typical example of system with
mass inversion phenomenon.
Recently, Lei et al. (2013a, b) also used this kind of
wind to explain the HB morphology 
in GCs, but they did not consider the process of
late hot flash and its effects on the formation of BHk stars.

Since the torque due to tidal friction depends on $(R/R_{\rm{L}})^{6}$
($R$ is the radius of the primary, while $R_{\rm{L}}$ is the Roche lobe radius of the primary. 
Zahn 1975; Campbell \& Papaloizou, 1983 ), Tout \& Eggleton
(1988) used the following equation to describe the
tidally enhanced stellar wind of the red giant primary:
\begin{equation}
\dot{M}=-\eta4\times10^{-13}(RL/M)\{1+B\rm_{w}\times \rm min[{\it(R/R\rm_{L})}^{6},
\rm 1/2^{6}]\},
\end{equation}
where $\eta$ is the Reimers mass-loss efficiency (Reimers 1975),
and $B\rm_{w}$ is the efficiency of the tidal
enhancement for the stellar wind.  Here $R$, $L$, and $M$ are
the radius, luminosity and mass of the primary in solar units.
Tout \& Eggleton introduced a saturation
for $R \ge 1/2R\rm_{L}$ in the above expression, because it is expected
that the binary system is in complete corotation for $R/R\rm_{L} \ge
0.5$. In this study, we set $\eta$ = 0.45 (Renzini \& Fusi Pecci 1988;
also see the discussion in Sect 4.2 of Lei et al. 2013b), $B\rm_{w}$ = 10000
(e.g., a typical value used in Tout \& Eggleton 1988).

To study the contribution of tidally enhanced stellar wind
on the formation of BHk stars, equation (1) was added into the
detailed stellar evolution code, Modules for
Experiments in Stellar Astrophysics (MESA;
Paxton et al. 2011, 2013).
MESA provides a variety of up-to-date physics modules,
and it can evolve low mass stars through the helium
core flash phase with details. Furthermore, it offers the
capacity that users can create zero age main-sequence
(ZAMS) and zero age horizontal branch (ZAHB) models
with different chemical compositions conveniently.
We evolve binary systems using the binary module of MESA,
version 6208. In the study, all the default values
for input physics in MESA are used except for the opacity tables.
OPAL type I opacity tables are the default ones for MESA, but
OPAL type II opacity tables allow time dependent variation of
C and O abundances independent of initial metal distribution,
thus it gives a better approximation of
the appropriate opacities during and after helium burning.
Therefore, OPAL type II opacity tables are used in this study. 

In the calculation, the secondary stars are assumed
to be MS stars and the mass ratio of primary
to secondary is set to 1.6 (see the discussion in Sect 4 of
Lei et al. 2013a for different values of $q$). Since what we concern is
the primary star which may become a BHk star after undergoing huge
mass loss on the RGB, we do not evolve the secondary star
during the binary evolution. In our calculations, a
moderate metallicity, $Z$ = 0.001 which is a typical value for
GCs in our Galaxy, is adopted, and
a primordial helium abundance $Y$ = 0.24 is adopted.
In our models, the mass of primary star at ZAMS is 0.83$M_{\odot}$
which corresponds to an age of about
12 Gyr at the RGB tip, and it is consistent with the typical age of GCs. 
We present in Table 1 the main input parameters described above.
From left to right, it gives the mass of
primary star at ZAMS, metallicity, helium abundance,
mass ratio of primary to secondary, Reimers
mass loss efficiency and tidal enhancement efficiency, respectively.

\begin{table*}
 \begin{minipage}{100mm}
  \caption{Main input parameters used in the study.}
  \begin{tabular}{@{}ccccccc@{}}
  \hline
  $M_{\rm ZAMS} (M_{\odot})$    &  $Z$   &  $Y$   & $q$  & $\eta$  & $B\rm_{w}$  \\
  0.83\footnote{This value of mass corresponds to an age of about 12 Gyr at RGB tip with $Z$ = 0.001 and $Y$ = 0.24.}  &   0.001  & 0.24  &1.6 &  0.45  & 10000  \\
\hline
\end{tabular}
\end{minipage}
\end{table*}

When considering the tidally enhanced stellar wind
in binary evolution, the mass loss of the primary star
on the RGB depends on the orbital period (Lei et al. 2013a).
For a long initial orbital period, the primary may  
lose little envelope mass due to the weak effects of
the secondary and become a normal HB star after
helium core flash. On the other hand, for a moderate short
period, the primary may lose much or nearly the whole envelope
mass and become a canonical EHB or BHk star after
helium core flash. However,
if the orbital period is short enough, the primary may
lose too much envelope mass on the RGB to ignite
helium in its core, and it will die as an helium white dwarf.
For this reason, four different initial orbital periods are used
in our calculation, which are 2200, 2000, 1610 and 1600 in days,
respectively. For each period, we evolve the primary star
from ZAMS to post HB phase and obtain the evolution parameters of
the primary. The results are given in next section.

\section{results}

In this section, we present the evolution for the primary stars
from ZAMS to post HB stage after undergoing huge mass loss
on the RGB. To compare our results with the observations in the CMD of GC,
the effective temperatures and luminosity of the
stars at helium burning stage are transformed into colors and
magnitudes using the sellar spectra library compiled by
Lejeune et al. (1997, 1998).  We also compared
our results with the observation in
the $\rm{log}\it{g}-T\rm_{eff}$ diagram of NGC 2808.

\begin{figure}
\centering
\includegraphics[width=87mm]{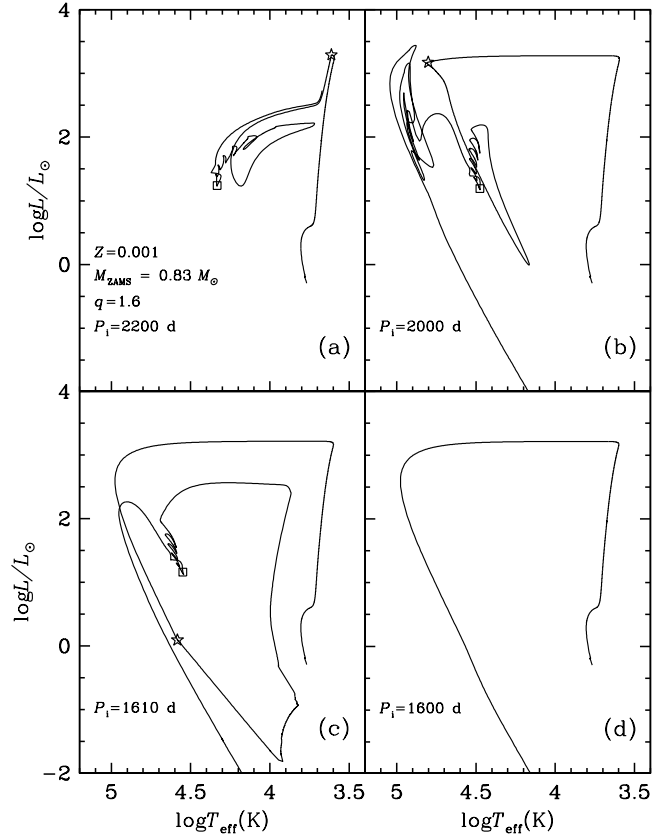}
\begin{minipage}[]{90mm}
\caption{The evolution tracks for the primary star
undergoing tidally enhanced stellar wind. The
pentacles in each panel represent the position where
the primary helium core flash takes place, while the
open squares denote ZAHB position. The initial
orbital periods are labeled in each panel.}
\end{minipage}
\end{figure}

\subsection{Evolution tracks}

Fig. 1 gives the evolution tracks of a primary star with
an age of about 12 Gyr at RGB tip for different initial orbital
periods. The pentacles in this figure
denote the position where the primary helium core flash takes place,
while the open squares denote the ZAHB positions for each
evolution track. In Panel (a), with an initial orbital
period of $P_{\rm i}$ = 2200 d, the primary star undergoes a normal
helium core flash at RGB tip,  and it becomes a canonical
EHB star with $\rm{log}\it{T}\rm_{eff}$ = 4.33 and
$\rm{log}\it L/L_{\odot}$ = 1.24 at ZAHB after the helium core flash.
One can infer simply that if the orbital period
is larger than this one, the primary star may lose
less envelope mass and locate on redder and more luminous
position of HB (see Lei et al. 2013a).
For the track with $P_{\rm i}$ = 2000 d in Panel (b),
due to the huge mass loss, the primary star evolves off the RGB tip
and undergoes the helium core flash when crossing the H-R diagram.
This kind of flash is called the early hot flash
(Lanz et al. 2004; Miller Bertolami et al. 2008),
in which the flash mixing can not penetrate the envelope and change the
surface chemical abundance of the star. After the
helium core flash, the primary star for early hot flash locates
on a hotter and fainter HB position in H-R diagram with
$\rm{log}\it{T}\rm_{eff}$  = 4.48 and $\rm{log}\it L/L_{\odot}$ = 1.19.
However, for the track with
$P_{\rm i}$ = 1610 d in Panel (c), helium core flash of the primary
star occurs when the star is descending
the white dwarf cooling curve, which is called the late hot flash
(Lanz et al. 2004; Miller Bertolami et al. 2008).
In this case, hydrogen in the envelope can be mixed into helium burning interior
by the flash triggered mixing. Therefore, the surface of this kind of
star is helium and carbon enhanced (Cassisi et al. 2003), and the star
locates on the BHk position with highest temperature and faintest
luminosity in H-R diagram (e.g., $\rm{log}\it{T}\rm_{eff}$ = 4.55
and $\rm{log}\it L/L_{\odot}$ = 1.16).
In panel (d), with an initial orbital period of $P_{\rm i}$ = 1600 d, the
primary star fails to ignite helium in the core due to too much
mass loss on the RGB, and it dies as an helium white
dwarf. Obviously, any periods shorter than this one
would produce helium white dwarfs rather than HB stars due to
the too much mass loss on the RGB.\footnote{In our models, we do not consider
close binaries in which the primary may fill its Roche lobe on the RGB 
and transfer envelope to the secondary rapidly and form EHB stars
(see Han et al. 2002, 2003).}

\begin{table*}
 \centering
 \begin{minipage}{140mm}
  \caption{Evolution parameters at ZAHB for the tracks shown in Fig.1}
  \begin{tabular}{@{}llllllllll@{}}
  \hline
  $P_{\rm i}$ (days) & $M_{\rm ZAHB} (M_{\odot})$    & $M_{\rm core}
  (M_{\odot})$\footnote{The helium core mass is defined
  by the region in which the hydrogen mass fraction is
  lower than 0.01.}
  &  $\rm{log}\it{T}\rm_{eff}$   & $\rm{log}\it L/L_{\odot}$  & $\rm{log}\it{g}$
  & $X\rm_{surf}$  & $Y\rm_{surf}$ & $C\rm_{surf}$ & Flash status\\
    \hline
   2200\footnote{The minimum initial orbital period for the normal helium core flash at the RGB tip (see text for detail).} & 0.5036  & 0.4817  & 4.3323  & 1.2393  & 5.1833  & 0.7443  & 0.2531  & 1.6144$\times10^{-4}$ & normal \\
  2000\footnote{The minimum initial orbital period for early hot flash (see text for detail).}  & 0.4829  & 0.4814  & 4.4756  & 1.1941  & 5.7837  & 0.7443  & 0.2531  & 1.6161$\times10^{-4}$ & early \\
  1610 & 0.4702  & 0.4702  & 4.5498  & 1.1593  & 6.1035  & 0.0014  & 0.9701  & 1.0676$\times10^{-2}$ & late \\
  1600\footnote{Any initial periods shorter than this one would produce helium white dwarfs due to too much mass loss on the RGB.}   &--&--&--&--&--&--&--&--& WD \\
\hline

\end{tabular}
\end{minipage}
\end{table*}

We also give in Table 2 the evolution parameters at ZAHB for
the tracks shown in Fig. 1.  From left to right
of Table 2, it presents the initial
orbital period, stellar mass, helium core mass,
effective temperature, luminosity, gravity, surface hydrogen
mass fraction, surface helium mass fraction, 
surface carbon mass fraction and
the flash status (e.g., what kind of flash does the
primary star experience, 'normal' means the star
experiences a normal helium core flash at the RGB tip,
'early' means it experiences an early hot flash
when crossing the H-R diagram, 'late' means it
experiences a late hot flash when
descending the white dwarf cooling curve, while 'WD'
means the primary star fails to ignite helium and
dies as an helium white dwarf).
One can see from Table 2 that with the decreasing of
initial orbital periods for the binaries, more
envelope mass of the primary star is lost
due to tidally enhanced stellar wind.
Though the early hot flasher (e.g., $P_{\rm i}$ = 2000 d) shows
a little smaller helium core mass than the normal flasher
(e.g., 0.4814 $M_{\odot}$ vs 0.4817 $M_{\odot}$),
its envelope mass is much less than the normal one
(e.g., 0.0015 $M_{\odot}$ vs 0.0219 $M_{\odot}$).
Thus, it results in a difference of about 8400 K for the
ZAHB temperature between early hot flasher and normal
flasher. However, the mass fractions of hydrogen, helium
and carbon in the surface both for the
early hot flasher and the normal one are nearly the same,
which means that the chemical compositions in the
surface of the two models are not changed by the helium core flash process.
On the other hand, due to loss of
nearly the whole envelope mass on the RGB,
the late hot flasher (i.e., $P_{\rm i}$ = 1610 d)
presents the smallest helium core mass and the thinnest envelope
mass at ZAHB when comparing with the early hot flasher and the normal flasher.
As predicted by Brown et al. (2001), the late hot flasher in Table 2
shows hydrogen deficiency but helium and carbon
enhancement in the surface. 
Due to the highest helium abundance in the
surface (e.g., 0.97 by mass) and
thinnest envelope mass, the late hot flasher presents
highest effective temperature at ZAHB which is higher than
early hot flasher about 5500 K and is higher than
normal flasher for about 14000K.
Since the primary for $P_{\rm i}$ = 1600 d becomes
an helium white dwarf, we do not give the parameters
for this model in Table 2.

According to our calculations,
$P_{\rm i}$ = 2200 d is the minimum initial orbital period
for which the primary stars would experience
normal helium core flash at the RGB tip, while $P_{\rm i}$ = 2000 d
is the minimum initial orbital period for which the primary
stars would experience early hot flash. That is to say,
if the initial orbital periods
are longer than $P_{\rm i}$ = 2200 d, the primary stars would
experience normal helium core flash at the RGB tip and
become canonical HB stars (e.g., EHB, blue HB or red HB
stars, see Lei et al. 2013a), while if the initial orbital periods are shorter than
$P_{\rm i}$ = 2200 d but longer than $P_{\rm i}$ =  2000 d, the primary stars
would experience early hot flash and become normal EHB stars.
For 1600 d $< P_{\rm i} <$ 2000 d, the primary stars would undergo
late hot flash when descending the white dwarf cooling curve and become
BHk stars. However, if the initial orbital periods are shorter than $P_{\rm i}$
= 1600 d, the primary stars would fail to ignite helium in the core due to
too much mass loss on the RGB 
and become helium white dwarfs. 

Cassisi et al. (2003)
firstly presented the evolution of a metal-poor
low mass star from ZAMS to post HB phase through
a delayed helium core flash when the star is
descending the white dwarf cooling curve. In
their models, stars lose envelope mass on the RGB
through Reimers mass-loss law (Reimers 1975).
For the model undergoing late hot flash (e.g.,
$\eta$ = 0.60 in Cassisi et al. 2003),
the mass fraction of hydrogen, helium and carbon
in the surface is 4$\times 10^{-4}$, 0.96 and 0.029
respectively. For our model in Table 2 with an initial
orbital period of $P_{\rm i}$= 1610 d, which also experienced a
late hot flash, the surface hydrogen, helium and carbon mass
fraction are 1.4$\times 10^{-3}$, 0.97 and 0.011 respectively.
The slight difference for chemical abundance in
surface between the model in Cassisi et al. (2003)
and ours may be due to the different mixing efficiency
of the convective region used in the two
model calculations (also see the
comparison in Table 4 of Miller Bertolami et al. 2008
between their results and the ones in Cassisi et al. 2003).

\begin{figure}
\centering

   \includegraphics[width=80mm]{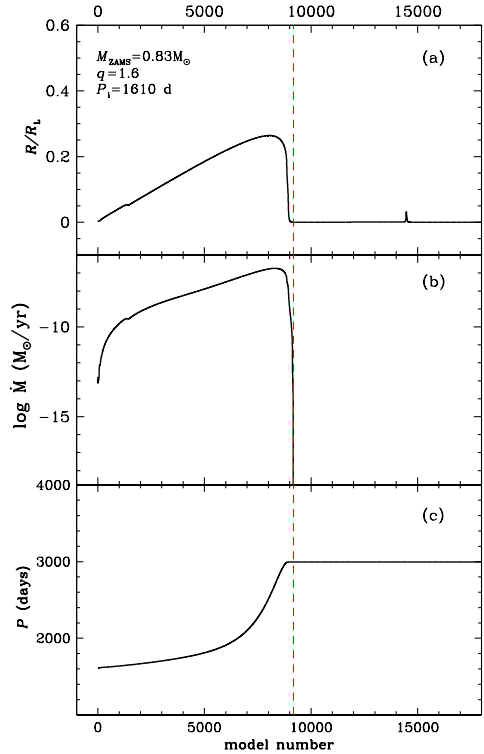}

\begin{minipage}[]{85mm}
 \caption{Binary evolution parameters vs model number. Model number
 is the time step calculated in MESA. The red-dashed line denotes the
 time when the primary helium core flash is taking place. Panel (a), (b) and (c)
 gives the evolution of the ratio between radius and
 Roche lobe radius, mass loss rate in stellar wind and
 orbital period, respectively.}
 \end{minipage}
\end{figure}

In figure 2, we give the evolution parameters
for one of our binary systems from ZAMS to
post HB, in which initial orbital
period $P_{\rm i}$ = 1610 d. The red-dashed line in Fig. 2
denotes the time when the primary helium core flash
is taking place.  Panel (a) shows the ratio of stellar radius to
Roche lobe radius for the primary star vs model
number\footnote{Model number is the time step calculated
in MESA and therefore not proportional to time.}.
One can see from Panel (a) that
the envelope of the primary star expands from
ZAMS and gradually closes to
the Roche lobe radius until
it reaches the RGB tip. According to equation (1),
mass loss from tidally enhanced stellar wind become more
and more stronger when the radius of the primary
star is being closer to its Roche lobe
radius. This process is apparently shown
in Panel (b), which gives the evolution of mass loss rate
in logarithms (i.e., {\rm log\it{\.{M}}} $M_{\odot}$/yr)
for the primary star. Due to
the huge mass loss on the RGB, the star 
contracts and evolves off the
RGB tip. At the same time, the mass loss
of the star decreases accordingly. When
the primary helium core flash takes place
on the white dwarf cooling curve, we shut down
the stellar wind of the star. One can see from
Panel (a), there is a small second expansion for
the radius of the star beyond the RGB tip.
This is due to the fact that the hydrogen in the envelope
was mixed into the inner region during the helium
core flash and burned into helium at high temperature,
which released large quantities of energy and push the
envelope towards lower temperatures.
Panel (c) in Fig. 2 shows the period evolution of the binary
vs model number. The orbital period of the binary
becomes more and more longer until the
star reaches the RGB tip (e.g.,
from 1610 d at ZAMS to about 3000 d at the RGB tip) due to the
mass loss and angular momentum loss from the
primary star. After the star
evolving off the RGB tip, mass loss
and angular momentum loss decrease largely,
thus it changes the orbital period a little.

\subsection{Comparison with observation for NGC 2808}

Brown et al. (2001) discovered the BHk
populations in the ultraviolet CMD of NGC 2808.
This cluster is one of the most massive globular
clusters in our Galaxy with an
intermediate metallicity of $Z$ = 0.0014 (i.e.,
[Fe/H] = -1.15, Harris 1996, version 2003) and an
age of 11 $\pm$ 0.38 Gyr (VandenBerg et al. 2013).
Since the parameters of GC NGC 2808 are similar with
the ones used in our model calculations, we
compare our results with this cluster in the
CMD and $\rm{log}\it{g}-T\rm_{eff}$ plane.
To compare our results with the observation in CMD, the effective
temperatures and luminosity of each evolution track shown in
Fig. 1 are transformed into colors and absolute magnitudes using
the sellar spectra library compiled by Lejeune et al. (1997, 1998).

\begin{figure}
\centering

   \includegraphics[width=85mm]{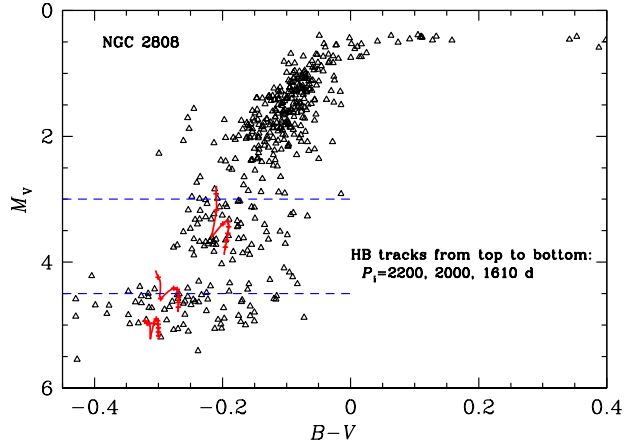}

\begin{minipage}[]{85mm}
 \caption{Comparing the evolution tracks at helium burning stage
shown in Fig. 1 with observation in the CMD of NGC 2808.
The mass of the primary at ZAMS is 0.83$M_{\odot}$,
the mass ratio of primary to secondary is 1.6, $Z$ = 0.001.
From top to bottom, the initial
orbital periods for the evolution tracks are $P_{\rm i}$ = 2200, 2000 and 1610 days.
The time interval between two adjacent
$+$ symbols in each track is $10^{7}$ yr.
The top blue-dashed line at $M_{\rm V}$
= $3^{m}$ roughly distinguishes canonical
EHB stars from classical blue HB stars, while
the bottom blue-dashed line at $M_{\rm V}$
= $4^{m}.5$ roughly distinguishes BHk stars from
canonical EHB stars (see the text for details).
The photometric data for NGC 2808 is from
Piotto et al. (2002).
} \end{minipage}
\end{figure}

Fig. 3 shows the comparison of our evolution tracks with
the observation in the CMD of NGC 2808. The
photometric data for this cluster is obtained by
Piotto et al. (2002) with the HST/WFPC2 camera in the
F439W and F555W bands. The magnitudes in F439W and F555W bands
are transformed into standard Johnson $B$ and
$V$ magnitudes by Piotto et al. (2002) using an
iterative procedure (see Sect 2.4 in their study). 
The distance module for the cluster
is set to 15.25 and the reddening correction $E(B-V)$ 
is set to 0.18 in Fig. 3 (Bedin et al. 2000;
D'Antona \& Caloi 2004). According to previous study for this cluster,
there are two apparent
gaps on the blue tail of NGC 2808 at $M_{\rm V}$
$\approx$ $3^{m}$ and $4^{m}.5$ (Bedin et al. 2000;
Moehler et al. 2004). The brighter gap distinguishes
canonical EHB stars from classical blue HB stars
(e.g., denoted by the top blue-dashed line in Fig. 3; Moeheler et al. 1997, 2000;
Momany et al. 2002), while the fainter gap separates
BHk stars from canonical EHB stars (e.g., denoted by the bottom
blue-dashed line in Fig. 3; Brown et al. 2001; Moehler et al. 2004).

In Fig. 3, open
triangles represent the observational HB stars in NGC 2808\footnote{
The red HB stars in this GC are not shown for clarity.},
while the three red-solid curves are the evolution
tracks at helium burning stage shown in Fig. 1.
From top to bottom, the initial orbital periods for the evolution tracks
are $P_{\rm i}$ = 2200, 2000, 1610 d respectively. 
The time interval between two adjacent
$+$ symbols in each track is $10^{7}$ yr.
For the top evolution track with $P_{\rm i}$ = 2200 d,
the primary star undergoes a normal helium core flash at the
RGB tip and locates on the canonical EHB regions
in the CMD (e.g., brighter than $M_{\rm V} \approx 4^{m}$ but
fainter than $M_{\rm V} \approx 3^{m}$).
The middle evolution track with $P_{\rm i}$ = 2000 d undergoes an
early hot flash and presents a little bit
smaller helium core mass but much thinner envelope mass 
than the top one (e.g., $P_{\rm i}$ = 2200 d, see Table 2).
Therefore, it occupies a bluer and fainter region than the
top evolution track in the CMD
(e.g., nearby the region for $M_{\rm V} \approx 4^{m}.5$).
Obviously, both the normal and the early hot flash track
cannot reach the faintest region in the CMD which is
occupied by BHk stars. On the other hand,
for the evolution track with $P_{\rm i}$ = 1610 d, it
experiences a late hot flash and show the smallest
helium core mass as well as highest helium abundance in the
surface and locates well on the BHk regions in
the CMD.

\subsection{$\rm{log}\it{g}-T\rm_{eff}$ plane}

\begin{figure}
\centering
   \includegraphics[width=87mm]{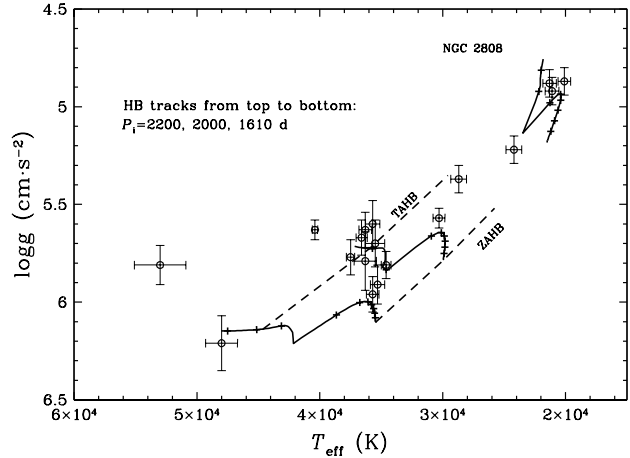}
   \begin{minipage}[]{90mm}
 \caption{Comparing the HB evolution tracks shown in Fig. 3 with
observation in the $\rm{log}\it{g}-T\rm_{eff}$ plane for NGC 2808.
The atmospheric parameters for BHk stars in NGC 2808 are
from Moehler et al. (2004).
The time interval between two adjacent
$+$ symbols in each track is $10^{7}$ yr.
The two dashed lines at bottom and top denote the
ZAHB and TAHB positions (see the text for details).}
\end{minipage}
\end{figure}

Fig. 4 shows the comparison between our evolution tracks and the observation
in $\rm{log}\it{g}-T\rm_{eff}$ plane for NGC 2808. In this figure, the open circles
denote the BHk and canonical EHB stars in NGC 2808 analyzed
by Moehler et al. (2004).
As the same with Fig. 3, the three solid curves in this figure
are the evolution tracks with initial orbital periods of
$P_{\rm i}$ = 2200, 2000, 1610 d from top to bottom, which
correspond to normal flasher, early hot flasher and late hot flasher respectively. 
The dashed line at the bottom denotes the ZAHB positions, while the
dashed line at the top denotes the terminal age HB (TAHB) positions
(TAHB is defined by the stage when central
helium abundance has dropped below 0.0001 in mass fraction).

Comparing with the evolution tracks, one
can see in Fig. 4 that the stars with $T\rm_{eff} \lesssim 32000$ K
are canonical EHB stars which are most likely
formed through normal or early hot helium flash.
On the other hand, most of the BHk stars with $T\rm_{eff} > 32000$ K
(Moehler et al. 2004; Moni bidin et al. 2012)
are well predicted by the late hot flash track for $P_{\rm i}$ = 1610 d.
As discussed in Moehler et al. (2004), the two
hottest stars with $T\rm_{eff} > 45000$ K may be
post HB stars which have exhausted
helium in the cores and evolve towards the white dwarf
cooling curve. One would be aware of that
the gravities for BHk stars obtained
by Moehler et al. (2004) seems to be generally lower than
the ones predicted by our late hot flash track. This discrepancy
between observation and model calculation also seems appear in
Fig. 5 of Moehler et al. (2004). A possible explanation
for this discrepancy is that the late hot flash scenario predicts
enrichment in carbon and nitrogen (Lanz et al. 2004),
while this enrichment is not considered in the model atmospheres
when obtaining the atmospheric parameters for
BHk stars in NGC 2808. However, Moehler et al. (2011)
obtained two sets of atmospheric parameters for the He-rich hot HB stars
in $\omega$ Cen with and without considering carbon and
nitrogen enrichment in the model atmospheres.
By comparing the two sets of gravities list in Table 4 of
Moehler et al. (2011), one can infer that the gravities obtained
with C/N enhanced model atmospheres are generally larger
than the ones obtained without C/N enhanced model atmospheres
by a mean value of about 0.13 dex. Therefore, a better
match between the observation and model calculation would be
expected if the enrichment of carbon and nitrogen is
considered in the model atmospheres to obtain the atmospheric
parameters for BHk stars in NGC 2808 (also see the discussion in
Sect 4.4 of Moni Bidin et al. 2012).

\section{discussion}

Moehler et al. (2004) analyzed medium resolution
spectra for some BHk stars in NGC 2808 and 
obtained their atmospheric parameters, including
the surface helium abundance. Therefore, we
compare our results with the observation
in the log(${n\rm_{He}/\it{n}\rm_{H}}$) - $T\rm_{eff}$ plane of NGC 2808 in Fig. 5.
In this figure, open circles are BHk stars or canonical 
EHB stars in NGC 2808 analyzed by Moehler et al. (2004),
while asterisk, open triangle
and cross denote our ZAHB models for normal flasher, early hot flasher and
late hot flasher respectively.
As seen in Fig. 5, our ZAHB models which undergo
normal flash and early hot flash (e.g., asterisk
and open triangle) show higher helium
abundance than the stars in NGC 2808 with similar
effective temperatures (e.g., in the range of 20000 K $\sim$ 32000 K).
As discussed in Sect 3, these stars are most likely
the canonical EHB stars and may have normal chemical
composition in the surface but undergo gravitational settling and
radiative levitation (Grundahl et al. 1999; Moehler et al. 2004;
Miller Bertolami et al. 2008) which may lead to
a helium poor and hydrogen rich surface. However, these processes are not
considered in our models, and it may causes the
discrepancy in helium abundance between our models
and the observation shown in Fig. 5.
Except for the two hottest stars in Fig. 5
(e.g., $T\rm_{eff}$ $>$ 45000 K), all the BHk stars in NGC 2808 with
$T\rm_{eff}$ $>$ 32000 K show solar or super solar helium
abundance\footnote{In Fig.5, log(${n\rm_{He}/\it{n}\rm_{H}}$) = -1 denotes
the solar helium abundance.}. This result is
consistent with the late hot flash scenario which
predicts a helium enhancement in the surface.
However, our ZAHB model for late
hot flash shows a value of log(${n\rm_{He}/\it{n}\rm_{H}}$) $\approx$ 2.23 
(e.g., the cross in Fig. 5), which is higher than 
the most He-rich BHk star in NGC 2808
(e.g., log(${n\rm_{He}/\it{n}\rm_{H}}$) $\approx$ 1) by about 1.2 dex.
A similar discrepancy also appeared in Cassisi et al. (2003)
when they compared their results with
the BHk stars observed in $\omega$ Cen. A value of
log(${n\rm_{He}/\it{n}\rm_{H}}$) $\approx$ 2.8 is
obtained for the late hot flash model
in Cassisi et al. (2003), while
the highest helium abundance for BHk stars
in $\omega$ Cen obtained by Moehler et al. (2002)
is log(${n\rm_{He}/\it{n}\rm_{H}}$) $\approx$ 0.94. 
Cassisi et al. (2003) suggested that a reduction by a factor of
2 $\times 10^{4}$ for mixing efficiency would
increase the residual of hydrogen by an order
of magnitude. Similarly, this explanation is a possible solution
for the discrepancy presented here, which means
that the flash mixing efficiency may be overestimated
in our model calculation.
On the other hand, Moehler et al. (2004) proposed
another possible solution that this discrepancy
could be explained if some hydrogen survived the
flash mixing and late diffused outward into the
surface. Miller Bertolami et al. (2008) estimated that 
the time scale of such a diffusive processes is
about $10^{6}$ yr, which is much shorter than the
HB life time of about $10^{8}$ yr.

The late hot flash scenario also predicts a carbon
enhancement in the surface of BHk stars (Brown et al. 2001). 
In the study of Moehler et al. (2007), the surface abundance for 
helium, carbon and nitrogen were obtained for BHk stars and
canonical EHB stars in $\omega$ Cen. They found that
carbon enrichment strongly correlated with helium
enrichment, e.g., with carbon enhancement of at least 1\% up to 3\% by mass
for the He-rich stars (see Fig. 3 in Moehler et al. 2007). 
Moreover, Latour et al. (2014) obtained the spectra of 38 hot EHB 
stars ($T\rm_{eff}$ $>$ 30000 K) in $\omega$ Cen using the MXU mode of FORS2/VLT 
and determined their fundamental parameters 
(e.g., effective temperature, gravity, surface abundance for 
hydrogen, helium and carbon). 
They clearly found that there is a positive correlation between carbon and 
helium enrichment for their He-rich EHB stars (see Fig. 7 in Latour et al. 2014),  
and the surface carbon mass fraction is up to 
about 1.5\% for the most He-rich EHB stars.  
These observations are consistent with the 
results predicted by our late hot flash model in this study  
(e.g., in Table 2, the surface carbon for the late hot flash model 
is about 1.1\% by mass, which is nearly two orders of magnitude higher 
than the value for normal flash and early hot flash model), and it seems to favor the 
late hot flash scenario for the formation of He-rich EHB stars 
in $\omega$ Cen (but also see the discussion in Sect 4.2 of Latour et al. 2014).

\begin{figure}
\centering

   \includegraphics[width=85mm]{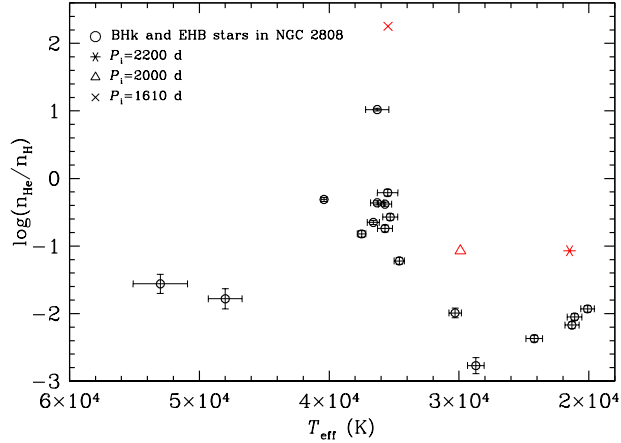}

\begin{minipage}[]{85mm}
 \caption{Comparing helium abundance between
 our models at ZAHB and the BHk stars of NGC 2808
 in log(${n\rm_{He}/\it{n}\rm_{H}}$) - $T\rm_{eff}$ plane. The
 observation data is from Moehler et al. (2004).}
 \end{minipage}
\end{figure}

BHk stars have been so far found in a few most massive
GCs. Dieball et al. (2009) investigated
the relationship between the physical or structural parameters
of GCs and the presence of BHk stars.
They found that the mass of GCs is the
most important fact linked to the formation
of BHk stars when comparing
the GCs with and without BHk stars.
A possible solution for this result is
that more massive GCs host more binaries, and
the binary interaction (e.g., tidally enhanced
stellar wind) may provide the huge mass loss
on the RGB which is needed for the late hot flash
scenario to form BHk stars. From Table 1 of
Dieball et al. (2009), the GCs hosting BHk stars
also show long relaxation times, large core
radii, small concentration and large escape
velocities, which means these GCs are not
core-collapsed. The most likely explanation is that
a significant fraction of binaries exist in these GCs
and decelerate or prevent the core collapse process.

Recently, low close binary fraction was found for
EHB stars in GCs (Moni Bidin et al. 2008, 2011), which
demonstrated a different formation channel between
EHB stars and their counterparts in the field,
subdwarf B (sdB) stars (Han et al. 2002, 2003).
This result does not contradict with ours.
One can see from Table 2 in this
study, the initial orbital periods of the binaries
which can produce BHk stars or canonical EHB stars
by tidally enhanced stellar
wind are longer than 1000 d. Moreover,
the orbital periods of the binary become even longer
during the binary evolution (see Panel (c) in Fig. 2).
Thus, it is difficult for these systems to
be identified as binaries from observation.
Radial velocity (RV) survey is likely
unable to find these binaries as well due to their long
orbital periods, since the RV variations for these
systems would be too small. However, some of
these binaries may be detected by eclipses, though
long campaigns would be required.
In our models, the secondary is a MS star which is not too
faint (e.g., 0.52$M_{\odot}$ for $q$ = 1.6). If this is the
case, the BHk primary star could be redder and more luminous than
as a single hot HB star in the optical CMD of GCs. In fact,
Castellani et al. (2006) found some peculiar HB stars
(named HBp stars in their study)
which are cooler or redder than the typical hot HB stars in
the optical CMD of NGC 2808 (see Fig. 2 in Castellani et al. 2006),
and they suggested that at least a fraction of these stars are a binary
nature with the faint companions having not been detected.
On the other hand, Bedin et al. (2000) and Iannicola et al. (2009)
found no distinct radial distribution differences for HB stars
(including EHB stars) in NGC 2808,
which argues against the binary nature for these stars.
Since binary systems would be segregated into the
core of the GC due to their massive masses,
the BHk stars would be more central concentrated if they are
formed through binaries. One
roughly possible explanation for this contradiction is that
the BHk primary stars may be diffused out the GC's core by
exchange interactions during the evolution of the cluster.
Due to the high density in the core of GCs, there are short-lived
three and four-body gravitational encounters where a star may be exchanged
into an existing binary and displaces one of the components of that binary (Heggie 1975).
By doing N-body simulations for 100,000 stars in cluster, Hurley
et al. (2007) found that the primordial binaries can be
replaced by new dynamical or exchange binaries through exchange interaction,
and these exchange binaries even become a dominate population in the core
of GC at the core-collapse phase (see Fig. 7 in their study). 

The tidal enhancement efficiency, $B\rm_{w}$, in this study is
set to 10000 which is a typical value used by Tout \& Eggleton (1988), but
a different value of $B\rm_{w}$ will not influence much on the main results
obtained in this study (see the discussion in Lei et al. 2013a).
Since a smaller $B\rm_{w}$ would cause enough mass loss on the RGB and
produce BHk stars for a relative shorter period, while
a bigger one also would produce BHk stars with a relative longer period.
Though a fixed value for the mass ratio of
primary to secondary is used in this study (e.g., $q$ = 1.6),
as discussed in Lei et al. (2013a), we found limited
effects for different values of $q$ on
the mass loss of the primary star, and
it just alters a little bit of the initial binary period
which is needed to produce BHk stars.
Finally, it is worth mentioning that, though the
parameters used in this study are suitable for GCs (e.g., age, metallicity),
the scenario proposed here (i.e., the tidally enhanced stellar wind)
is also a possible explanation for the formation of hot subdwarf stars in the field, which
has been previously suggested by Han et al. (2010), but the detailed
investigation for this field is out the scope of current study. 

\section{Conclusions}

In this paper, we suggested that
tidally enhanced stellar wind
in binary evolution may provide the huge mass
loss on the RGB which is needed for
the late hot flash scenario to explain the
formation of BHk stars. The tidally
enhanced stellar wind was added into
detailed stellar evolution code, MESA, to
investigate its contributions on the formation of BHk stars. 
Four different initial orbital periods for binaries are adopted
in the calculations, e.g., $P_{\rm i}$ = 2200, 2000, 1610 and
1600 days. We found that, if the initial orbital period are longer
than $P_{\rm i}$ = 2200 d, the primary stars would
experience normal helium flash at the RGB tip and become 
canonical HB stars, e.g., EHB,
blue HB or red HB stars, while if the 
initial periods are shorter than $P_{\rm i}$ = 2200 d but longer than $P_{\rm i}$ = 2000 d,
the primary stars would experience early hot flash and become
canonical EHB stars. For 1600 d $< P_{\rm i} <$ 2000 d, the primary stars
undergo late hot flash when descending the white dwarf cooling curve and
become BHk stars after helium core flash. However,
if the initial orbital periods are shorter than $P_{\rm i}$ = 1600 d,
the primary stars would become helium white dwarfs due to too much mass loss on the RGB. 
Our results are consistent
with the observation in the CMD
and the $\rm{log}\it{g}-\it{T}\rm_{eff}$ plane for NGC 2808.
However, the helium abundance of our
models seems to be higher than the ones observed
in BHk stars. It may be due to the fact that
gravitational settling and radiative levitation
are not considered in our models as well as
the fact that flash mixing efficiency is overestimated
in the calculation. The orbital periods
for the binary systems that can produce BHk stars or
canonical EHB stars by tidally enhanced stellar wind may be
too long to make these systems identified as
binaries in observation. Our results suggested that
tidally enhanced stellar wind in binary
evolution is a possible and reasonable formation
channel for BHk stars in GCs.

\section*{Acknowledgments}

It is a pleasure to thank the referee, Christian Moni Bidin, for his valuable suggestions
and comments, which improved the manuscript greatly. This work is supported by the Key Laboratory for
the Structure and Evolution of Celestial Objects, Chinese Academy of Science
(OP201302). Z. H. is partly supported by the Natural Science Foundation
of China (Grant No. 11390374),
the Science and Technology Innovation Talent Programme of
the Yunnan Province (Grant No. 2013HA005) and
the Chinese Academy of Sciences (Grant No. XDB09010202).

\label{lastpage}

\end{document}